\documentclass[doublecol]{epl2} 

\usepackage{amsmath}

\newcommand{\be}{\begin{equation}}
\newcommand{\ee}{\end{equation}}

\newcommand{\bean}{\begin{eqnarray}}          
\newcommand{\eean}{\end{eqnarray}}

\newcommand{\bea}{\begin{eqnarray*}}          
\newcommand{\eea}{\end{eqnarray*}}

\def\p{\partial}
\def\d{{\nabla}}

\def\Int{\int\limits}

\def\Eq#1{eq.~(\ref{#1})}
\def\art#1{[\ref{#1}]}

\title{Flat Universe with Hyperbolic Voids}

\author{V.G.~Gurzadyan\inst{1,2} \and A.A.~Kocharyan\inst{1,2,3}}
\shortauthor{V.G.~Gurzadyan \and A.A.~Kocharyan}

\institute{                    
  \inst{1} Yerevan Physics Institute, Armenia\\
  \inst{2} Yerevan State University, Armenia\\
  \inst{3} School of Mathematical Sciences, Monash University, Clayton, Australia
}

\abstract{The properties of geodesics flow are studied in a Friedmann-Robertson-Walker metric perturbed due to the inhomogeneities of matter. The basic, averaged Jacobi equation is derived, which reveals that the low density regions (voids) are able to induce hyperbolicity, even if the global curvature of the Universe is zero or slightly positive. It is shown that the energy independence is a characteristic property of these geometric effects. The importance of these conclusions is determined by the temperature independent ellipticity of excursion sets and regions of different randomness found in Kolmogorov CMB maps.  
}

\begin{document}

\maketitle

\section{Introduction}

The increase in the accuracy of Cosmic Microwave Background (CMB) radiation measurements has 
not only largely revealed the set of the basic cosmological parameters \art{cmb}, but has also stipulated the importance of the search of non-gaussian effects. The list of already 
reported non-gaussianities includes the low multipole alignment anomaly, the Cold Spot, etc (e.g. \art{anomaly} and references therein). 
The CMB distortions due to the cumulative contribution of inhomogeneous matter distribution in the Universe 
are among the discussed effects, which on the one hand, can influence the primordial non-gaussianities, on the other hand, will enable the link with the observed filaments (e.g. \art{lense}). 

Another empirical fact for potential interest here is the temperature independent ellipticity of anisotropies found in CMB sky maps \art{ell}.  For example, it was noticed \art{Wiltshire} that, if the nature of the dark energy is due to the matter inhomogeneities, then it might be linked with those ellipticities in the CMB maps. For further consideration
of the backreaction of the inhomogeneities see \art{B}.

Here we consider within a geometric framework the CMB distortions both due to the matter distribution 
and the overall geometry. This approach is based on the concept of the instability of 
geodesic flows and involves the Hamiltonian formalism, known by its 
efficiency in the problems of instability of dynamical systems \art{Arnold}.  

It is shown that the temperature independence of the distortions of CMB excursion sets is a generic feature 
of the geometric nature of the effect, as studied initially in \art{GK}. 
Namely, whatever is the agent causing the hyperbolicity itself, one has an empirical indication 
of a geometric effect, thus supporting such a nature for temperature threshold 
independence of ellipticities \art{ell}. 

Then, we show that the low density regions, voids, can induce hyperbolicity even in conditions of globally flat or positively curved Universe.  

To have possibly self-consistent account, we start with a brief account of the reduction of the
null geodesics from (3+1) to 3-spaces, then move to the averaged Jacobi equation and  
the conditions for the instability of the flows.

\section{Anosov flows}

The basic steps can be separated as: (a) to find out the conditions when the geodesics in (3+1)D can
be projected as geodesics in 3D Riemannian manifold, (b) the Hamiltonian representation of the flows,
(c) the time correlations functions of Anosov flows.  

Let $M$ be a smooth $d$ dimensional manifold with a smooth Riemannian metric $g$,
$\tau: TM\to M$ be the tangent bundle over $M$, and $K: TE\to E$ the linear connection on a vector bundle $\pi: E\to M$ over $M$, such that for each $m\in M$ and $X\in E_m=\pi^{-1}(m)$ the restriction
$K: T_XE\to E_m$ is linear and surjective. Then, the following diagram is commutative
$$
  \begin{array}{rcl}
             TE  & \stackrel{\textstyle K}{\longrightarrow}& E\\
\tau_E\downarrow &                                         & \downarrow\pi\\
              E  & \stackrel{\textstyle\pi}{\longrightarrow}& M.
   \end{array}
$$
For $X\in E_m$ we define
$$
  Ver(X)=\ker(T\pi: T_XE\to T_mM),
$$
and
$$
  Hor(X)=\ker(K_X: T_XE\to E_m),
$$
to be the vertical and horizontal subspace of $T_XE$, respectively. Therefore,  
$$
  T_XE=Hor(X)\oplus Ver(X).
$$

The covariant derivative $\nabla$ of a differentiable section $\xi: M\to E$ is defined by 
$\nabla \xi=K\circ T\xi$. For any smooth curve $c:R\to M:t\mapsto c(t)$ with tangent vector $\dot{c}=Tc.\frac{d}{dt}=u$ and tensor field $\xi$ along the curve $c$ we have
$$
  \d_u\xi=\d_{\dot{c}}\xi=K.\dot{\xi}=K\circ T\xi.\frac{d}{dt}.
$$

The isomorphism between $T_XTM$ and
$T_{\tau\circ X}M\oplus T_{\tau\circ X}M$ can be stated as
$$
  T\tau\times K: T_XTM\to T_{\tau\circ X}M\oplus T_{\tau\circ X}M: 
                Y\mapsto(T\tau.Y, K.Y).
$$

It is well known that on a smooth Riemannian manifold there exists exactly one torsion-free Riemannian connection, the Levi-Civita connection. Thus, for the Levi-Civita connection the Ricci identity holds $\nabla[g(X,Y)]=g(\nabla X,Y)+g(X,\nabla Y)$, and it is torsion-free, i.e. for any vector fields $X$ and $Y$ we have $\nabla_XY-\nabla_YX=[X,Y]$. The Riemannian tensor is given in terms of a covariant derivative by the following formula
$$
  Riem(X,Y)Z=([\nabla_X,\nabla_Y]-\nabla_{[X,Y]})Z.
$$
The linear transformation $\Re_X:Y\to Riem(Y,X)X$ is known as the curvature transformation.

Let $S$ be a smooth vector field on the tangent bundle $TM$
of a smooth Riemannian manifold $M$
$$
  S:TM\to TTM: u\mapsto(u,F)\in Hor(\tau\circ u)\oplus Ver(\tau\circ u).
$$
We denote by $u:R\to TM\ : t\mapsto u(t)=f^tu_0$ an integral curve of the vector field 
$S$ passing through the initial point $u_0\in TM$ i.e. $\dot{u}=S(u)$, and $u(0)=u_0$.
The group of diffeomorphisms $\{f^t\}$, $t\in R$
$$
  f^t: TM\to TM\ :u_0\mapsto u(t),
$$
is a one-parametric dynamical system or flow of the vector field $S$ on $TM$.
The curve $u(t)$ is the flow line or integral curve
starting at $u(0)=u_0$, $c=\tau\circ u$ is a trajectory.

If $u(t)$ is an integral curve of the vector field $S$
starting at $u_0$, then $c=\tau\circ u$ is a solution of the
following equation
$$
          \d_{\dot{c}}\dot{c}=F,
$$
determined by $c(0)=\tau u_0$ and $\dot{c}(0)=u_0$. In addition, if $\d_{\dot{c}}\dot{c}=F$, then
$\dot{c}(t)=f^t\dot{c}(0)$.

We will discuss vector fields $S$, such that $F=0$. In order to study chaotic properties of dynamical systems we need to investigate solutions of the equation for invariant vector fields $Z$ along the curve $c$, i.e. $Z(t)=Tf^tZ(0)$. They are in $1:1$ correspondence with the solutions of the Jacobi equation
\begin{equation}
  \d^2_uz+\Re_u(z)=0. \label{Jacobi}
\end{equation}
The correspondence is given by
$$
  z(t)\leftrightarrow Z(t)=(z(t),\d_uz(t))\in Hor(u(t))\oplus Ver(u(t)).
$$

It can be proved that the dynamical system defined by a vector field $S(u)=(u,0)$ is a geodesic flow with Lagrangian $L(x,u)=\tfrac{1}{2} g(u,u)$
\begin{equation}
\begin{cases}
\dot{c}=u,\cr
\d_uu=0,\cr
\d^2_uz+\Re_u(z)=0,\ 
\end{cases} \label{d0}
\end{equation}
and $g(u,u)=1$. The Jacobi field $z$ (solution of the Jacobi equation) is said to be
orthogonal to $u$, if $g(u,z)=0$. Any Jacobi field has the following form $z=n+const\cdot u$, 
where $n$ is orthogonal to $u$, i.e. $g(u,n)=0=g(u,\d_un)$. Thus, $n$ is orthogonal to $u$ for all $t$, if is orthogonal at $t=0$. It is a very important property of the geodesic flow, for it means that equation for $n$ is the same as for the $z$.

One can derive an equation for the length of $n$ from the Jacobi equation
$$
 \d^2_un+\Re_u(n)=0.
$$
Let $n=\ell\hat{n}$, where $g(n,n)=\ell^2$, $g(\hat{n},\hat{n})=1$, $g(\hat{n},u)=0$, then we get
\begin{equation}
  \ddot{\ell}+[K(u,\hat{n})-g(\d_u\hat{n},\d_u\hat{n})]\ell=0,\label{lOriginal}
\end{equation}
where the sectional curvature has the following form
$$
 K(u,\hat{n})=g(\Re_u(\hat{n}),\hat{n})=g(\Re(\hat{n},u)u,\hat{n}).
$$

If $M$ is compact and $K(X,Y)<K_0<0$ for all orthonormal vectors $X$ and $Y$ at any point of $M$, then the geodesic flow is an Anosov system \art{Anosov}. 
This condition is rarely fulfilled for physical systems.
Nevertheless, considering below the perturbed geodesic flow and to outline the link with basics of Anosov systems, we will refer as of Anosov instability, even though the inequalities are fulfilled for averaged quantities only.
\Eq{lOriginal} is usually very difficult to solve, so one is forced to find a simpler equation, the solutions of which approximate the original one. Substituting 
$$
  \frac{1}{d-1}\left(g-\frac{u\otimes u}{g(u,u)}\right)
$$
for $\hat{n}\otimes\hat{n}$ in $K(u,\hat{n})$ we get the following averaged equation
\begin{equation}
  \ddot{\ell}+\tfrac{1}{d-1}\ Ric(u,u)\ \ell=0,\label{lAverage1}
\end{equation}
where we have dropped the non-positive term $-g(\d_u\hat{n},\d_u\hat{n})\le0$ from \Eq{lOriginal}.
One may average this again by replacing $u\otimes u$ with $\frac{1}{d}g$ 
to get
\begin{equation}
  \ddot{\ell}+\tfrac{1}{d(d-1)}\ R\ \ell=0\ \label{meanl}\,
\end{equation}
(R is the scalar curvature). These operations are justified by the search of averaged properties of
geodesics in Robertson-Walker homogeneous and isotropic metric, and the fact that 
the solutions of the averaged equation are typically greater those of the
original equation \Eq{lOriginal} (cf. \art{Arnold}).

\section{Reduction of (3+1)D Null Geodesic Flow to 3D Geodesic Flow}

Let $V=M\times R^1$ be a (3+1)D smooth manifold  with the following metric
\begin{eqnarray}
ds^2 &=& g_{\mu\nu}dx^\mu dx^\nu \nonumber\\
&=& a^2(\eta)\ e^{2\psi(x)} 
\left[-d\eta^2 + e^{-2f(x)} d\sigma^2\right],\label{Metric}
\end{eqnarray}
where
\begin{eqnarray*}
d\sigma^2 = \gamma_{mn}(x)dx^mdx^n
\end{eqnarray*}
is any smooth metric on $M$; $\mu$, $\nu=0,1,2,3$; $m,n=1,2,3$. 

As is well known the geodesic flow on $V$ can be described by the following Hamiltonian
\begin{eqnarray*}
{\cal{H}}(p,x)&=&\tfrac{1}{2} g^{\mu\nu}(x) p_\mu p_\nu\\
&=&\tfrac{1}{2}\ a^{-2}(\eta)\ e^{-2\psi}
\left[-p_0^2 + e^{2f(x)} \gamma^{mn}p_mp_n\right]
\end{eqnarray*}
and Hamiltonian equations 
$$
\frac{dx^\mu}{ds} = \frac{\p {\cal{H}}}{\p p_\mu},\quad
\frac{dp_\mu}{ds} = -\frac{\p {\cal{H}}}{\p x^\mu}.
$$
Moreover, for the null geodesics we have ${\cal{H}}(p,x) =0$.

We are going to prove that projections of the geodesics from $V$ into $M$ will be geodesics on $M$ having the metric $\tilde\gamma=e^{-2f}\gamma$. It is easy to see that for the null geodesics
$$
\frac{\p {\cal{H}}}{\p x^0} 
= \frac{\p {\cal{H}}}{\p \eta}
= -2 \frac{\dot{a}}{a} {\cal{H}} = 0,
$$
therefore
$$
\frac{dp_0}{ds} = 0,
$$
and
$$
\frac{d\eta}{ds} =- a^{-2} e^{-2\psi} p_0,
$$
thus, $p_0$ is a constant ($E=-p_0$). For the ``space" coordinates we obtain
\begin{eqnarray*}
\frac{dx^m}{ds} &=& [a^{-2} e^{-2\psi}]\ \frac{\p H}{\p p_m},\\
\frac{dp_m}{ds} &=& -[a^{-2} e^{-2\psi}]\ \frac{\p H}{\p x^m},
\end{eqnarray*}
where
$$
H = \tfrac{1}{2}\tilde\gamma^{mn}p_mp_n = \tfrac{1}{2} e^{2f}\gamma^{mn}p_mp_n 
= \tfrac{1}{2} p_0^2.
$$
If we define $d\tau = a^{-2}e^{-2\psi}ds$, then we get
$$
\frac{dx^m}{d\tau} = \frac{\p H}{\p p_m},\quad
\frac{dp_m}{d\tau} = - \frac{\p H}{\p x^m},
$$
with $H = \tfrac{1}{2} p_0^2$. Thus, the $(3+1)$D null geodesic flow $(V,g,{\cal{H}},ds)$ is  reduced to a $3$D geodesic flow $(M,\tilde\gamma,H,d\tau)$. This generalises results of \art{Prigogine}.

\section{Energy dependence of the mixing}

It is obvious that
\begin{eqnarray*}
a^{-2}e^{-2\psi} 
= \frac{d\tau}{ds} 
= \frac{d\eta}{ds} \frac{d\tau}{d\eta}  
= \left[-a^{-2}e^{-2\psi} p_0\right]\ \frac{d\tau}{d\eta},
\end{eqnarray*}
hence $d\tau=-d\eta/p_0$. For $\hat{H}=H/p_0^2$ and $\hat\tau=p_0^2\tau$ we have
$$
\frac{dx^m}{d\hat\tau} = \frac{\p\hat H}{\p p_m},\quad
\frac{dp_m}{d\hat\tau} = - \frac{\p\hat H}{\p x^m},
$$
with 
$$
\hat H = \tfrac{1}{2} \hat{g}^{mn}\ p_mp_n = \tfrac{1}{2},
$$ 
where
$$
\hat{g}_{mn} = p_0^2\ \tilde\gamma_{mn}.
$$

From Ergodic theory (e.g. \art{ET}) we know that the rate of the decay of correlations (also called the rate of mixing)  is proportional to the $e^{-\chi\hat\tau}$, 
where Lyapunov exponent $\chi\sim \sqrt{R(\hat{g})}= |p_0|^{-1} \sqrt{R(\tilde\gamma)}$ and 
$\hat\tau = p_0^2\tau = |p_0|\eta$. Hence,

$$
\chi\hat\tau \sim \eta\sqrt{R(\tilde\gamma)}.
$$

Therefore, the mixing rate does not depend on the energy of photons ($E=-p_0$). {\it Thus, the relevant observable effects, such as the ellipticity of the excursion sets \art{GK}, have to be independent on photon's energy, temperature.}

\section{Instability in perturbed FRW Universe}

Let $V=M\times R^1$ be a (3+1)-D smooth manifold  with a perturbed Robertson-Walker metric of the form
\begin{eqnarray}
ds^2 &=& g_{\mu\nu}dx^\mu dx^\nu \nonumber\\
&=& -(1+2\phi)dt^2 + (1-2\phi)a^2(t)d\sigma^2,\label{PRWMetric}
\end{eqnarray}
where
$$
\gamma_{mn} = \left(1+\tfrac{k}{4}\left[(x^1)^2 +(x^2)^2 +(x^3)^2\right]\right)^{-2}\ \delta_{mn}
$$
is the metric of a 3-sphere ($k=1$), a 3-hyperboloid ($k=-1$), a flat 3-space ($k=0$), respectively.
As in \art{HolzWald} we assume that
$$
|\phi|\ll 1,
$$
and time derivatives of $\phi$ are much smaller than spacial derivatives, i.e.
$$
\left(\frac{\p\phi}{\p t}\right)^2 \ll a^{-2}\ \|\nabla\phi\|^2,
$$
and 
$$
\|\nabla\phi\|^2\ll |\Delta\phi|,
$$
where
$$
\|\nabla\phi\|^2 = \gamma^{mn}\ \frac{\p\phi}{\p x^m}\ \frac{\p\phi}{\p x^n},\ \  
\Delta\phi = -\nabla^2\phi.
$$

If we substitute $\psi=\phi$ and $f=2\phi$ and take into account that $e^{2\phi} \sim 1+2\phi$ and $e^{-2\phi} \sim 1-2\phi$, then the metric \Eq{Metric} will have the same form as \Eq{PRWMetric} for $d=3$, therefore we can use the above results.

To estimate the Lyapunov exponents one has to investigate the Riemann scalar curvature for the reduced space metric $\tilde\gamma=e^{-4\phi}\gamma$ (see \Eq{meanl}). It can be easily shown that

\begin{equation}\label{R}
R=6\ e^{4\phi}\left[k -\tfrac{4}{3}\left(\Delta\phi +\|\nabla\phi\|^2\right)\right]
\end{equation}
where  \art{HolzWald}
$$
-\Delta\phi = 4\pi a^2 \delta\rho.
$$
Therefore,
$$
\frac{R}{6} \simeq k -\frac{4}{3}\ \Delta\phi =k +\frac{16\pi}{3}\  a^2\delta\rho.
$$
Then, from
\begin{eqnarray*}
\bar\rho &=& \rho_0\ \frac{a_0^3}{a^3},\\
\left(\frac{\dot{a}}{a}\right)^2
&=& -\frac{k}{a^2} +\frac{\Lambda}{3}+\frac{8\pi\rho_0}{3}\ \frac{a_0^3}{a^3},\\
\frac{a}{a_0}&=&\frac{1}{1+z};\ \ \ H=\frac{\dot{a}}{a},\\ \\
\eta(t_0)-\eta(t)
&=&\int_{t}^{t_0}\ \frac{d\tau}{a(\tau)}
=  \frac{1}{a_0}\int_0^z\ \frac{d\xi}{H(\xi)}.
\end{eqnarray*}
one has
\begin{eqnarray*}
1&=& \Omega_k +\Omega_\Lambda +\Omega_m,\\
H^2&=&H_0^2\left[\Omega_\Lambda +[1- \Omega_\Lambda + \Omega_m z]\ (1+z)^2
\right],\\
\end{eqnarray*}
where
\begin{eqnarray*}
H_0^2&=& -\frac{k}{a_0^2} +\frac{\Lambda}{3}+\frac{8\pi\rho_0}{3},\\
\Omega_k &=& -\frac{k}{a_0^2 H_0^2},\ \ 
\Omega_\Lambda = \frac{\Lambda}{3H_0^2},\ \ 
\Omega_m = \frac{8\pi\rho_0}{3H_0^2}.
\end{eqnarray*}
Thus,
$$
\eta(t_0)-\eta(t)
=(a_0H_0)^{-1}\ \lambda(z,\Omega_\Lambda,\Omega_m),
$$
where
\begin{equation}
\lambda(z,\Omega_\Lambda,\Omega_m)
=\Int_0^z\frac{d\xi}{\sqrt{\Omega_\Lambda +[1- \Omega_\Lambda + \Omega_m \xi]\ (1+\xi)^2}}.
\end{equation}
And
\begin{eqnarray*}
\frac{R}{6}
&=& (a_0H_0)^2\left(-\Omega_k +2\delta_0\Omega_m\right)=(a_0H_0)^2\ r,
\end{eqnarray*}
where
$$
\delta_0\equiv\frac{\delta\rho_0}{\rho_0}.
$$
Finally, the \Eq{meanl} can be rewritten in the following form
\begin{equation}
\frac{d^2\ell}{d\lambda^2}+ r\ \ell = 0,
\end{equation}
where
\begin{equation}
r = -\Omega_k +2\delta_0\Omega_m.
\end{equation}
By definition $\delta_0\ge -1$, therefore, 
$r\ge -\Omega_k -2\Omega_m$.

These equations are the basic ones describing the mixing properties of null geodesics, i.e.
of the propagation of CMB both due the global geometry and local perturbations of metric (lensing). 
 
For example, when $\delta_0=0$, then $r=-\Omega_k$, as investigated in \art{GK}. 
While in the inhomogeneous Universe, the resulting effect can be different.

Indeed, the underdense regions with $\delta_0\le 0$, {\it the voids, contribute to the hyperbolicity}. 
Importantly, this conclusion does not require compactness of $M$. 

\section{Conclusions}

The geometric approach of theory of dynamical systems is used for the description
of the instability (mixing) of null geodesics peculiar to Anosov flows, both due to global curvature and locally perturbed metric in the context of the properties of CMB.
The corresponding equation, of averaged Jacobi form, and the Lyapunov exponents are derived, from which it follows that the low density regions, i.e. {\it the voids, would contribute to hyperbolicity, even if the Universe is globally flat or slightly positively curved.} 

The available observational parameters of the voids (i.e. of 30 Mpc size) obtained from the large scale galaxy redshift surveys
seem to support their role of hyperbolic lenses \art{GK08b}, namely, due to the cumulative effect of such
voids (parametric resonance), and not due to a single void of the horizon scale. The hyperbolicity of the voids and hence the loss of correlations in the propagating photon beams can be measured by the 
assigned degree of randomness in the CMB maps \art{GK08c}.
      
It is also shown that the peculiar property of the geometric mixing is its independence on the energy of photons. This fact is in agreement with the temperature independent ellipticity of anisotropies found in the CMB sky maps, first of COBE, then of Boomerang and WMAP \art{ell}, thus supporting the geometric nature of the effect.  

This approach can be efficient also at numerical simulations of CMB sky maps 
cross-correlated with the distributions of the filaments.

\acknowledgments
We are thankful to C.Misner, R.Kerr and D.Wiltshire for valuable comments.


\begin{thebibliography}{0}

\bibitem{cmb}
	\label{cmb}
	\Name{de Bernardis P. {\it et al.}}
	\REVIEW{Nature}{404}{2000}{955}.
	\Name{Spergel D. {\it et al.}}
	\REVIEW{ApJS}{170}{2007}{377}.
	\Name{Komatsu E. {\it et al.}}
	arXiv:0803.0547.


\bibitem{anomaly}
\label{anomaly}
	\Name{Rakic A., Schwarz D.J.}
	\REVIEW{Phys.Rev.}{D75}{2007}{103002}. 
	\Name{Copi C.J. {\it et al.}}
	\REVIEW{Phys.Rev.}{D75}{2007}{023507}
	\Name{Gurzadyan V.G., Starobinsky A.A. {\it et al.}}
	\REVIEW{A \& A}{490}{2008}{929}.     


\bibitem{lense}
\label{lense}
	\Name{Lewis A., Challinor A.}
	\REVIEW{Phys.Rept.}{429}{2006}{1}.
	\Name{Slosar A., Hirata C. {\it et al.}}
	arXiv:0805.3580.
	\Name{Granett B.R., Neyrinck M.C., Szapudi I.}
	arXiv:0805.3695.


\bibitem{ell}
\label{ell}
	\Name{Gurzadyan V.G., Torres S.}
	\REVIEW{A \& A}{321}{1997}{19}.
	\Name{Gurzadyan V.G., de Bernardis P. {\it et al.}}
	\REVIEW{Mod.Phys.Lett.}{A20}{2005}{893}.
	\Name{Gurzadyan V.G., Bianco C.L. {\it et al.}}
	\REVIEW{Phys.Lett. A}{363}{2007}{121}.


\bibitem{Wiltshire}
\label{Wiltshire}
	\Name{Wiltshire D.L.}
	\REVIEW{New J.Phys}{9}{2007}{377}. 
	\Name{Leith B.M., Cindy Ng S.C., Wiltshire D.L.}
	\REVIEW{ApJ}{672}{2008}{L91}. 

\bibitem{B}
\label{B}
	\Name{Zalaletdinov R.M}
	\REVIEW{Gen.Rev.Grav.}{24}{1992}{1015}. 
	\Name{Buchert T.}
	\REVIEW{Gen.Rel.Grav.}{40}{2008}{467}.
	\Name{Buchert T., Carfora M.}
	\REVIEW{Class.Quant.Grav.}{25}{2008}{195001}.
	\Name{Larena J., Alimi J.-M. {\it et al.}}
	arXiv:0808.1161.
 

\bibitem{Arnold}
\label{Arnold}
	\Name{Arnold V.I.}
	\Book{Mathematical Methods of Classical Mechanics}
	\Publ{Springer-Verlag}
	\Year{1978}.

\bibitem{08}
\label{GK}
\Name{Gurzadyan V.G., Kocharyan A.A.}
\REVIEW{A \& A}{260}{1992}{14};
\REVIEW{Europhys. Lett.}{22}{1993}{231}.

\bibitem{Anosov}
\label{Anosov}
	\Name{Anosov D.V.}
	\REVIEW{Comm. Steklov Mathematical Inst.}{90}{1967}.

\bibitem{ET}
\label{ET}
	\Book{Handbook of Dynamical Systems}
	\Editor{Hasselblatt B., Katok A.}
	\Publ{North Holland}
	\Year{2002}.   

\bibitem{Prigogine}
\label{Prigogine}
	\Name{Lockhart C.M., Misra B. and Prigogine I.}
	\REVIEW{Phys. Rev.}{D25}{1982}{921}.

\bibitem{HolzWald}
\label{HolzWald}
	\Name{Holz D.E., Wald R.M.}
	\REVIEW{Phys. Rev.}{D58}{1998}{063501}.

\bibitem{GK08b}
\label{GK08b}
	\Name{Gurzadyan V.G., Kocharyan A.A.}
	\REVIEW{A \& A}{493}{2009}{L61}.

\bibitem{14}
\label{GK08c}
	\Name{Gurzadyan V.G., Kocharyan A.A.}
	\REVIEW{A \& A}{492}{2008}{L33}.
	\Name{Gurzadyan V.G., Allahverdyan A.E. {\it et al}}
	\REVIEW{A \& A}{{\it in press}}{2008}, arXiv:0811.2732.

\end{thebibliography}
\end{document}